\documentclass[aps,prl,twocolumn,superscriptaddress,preprintnumbers,amsmath,amssymb,floatfix,nofootinbib]{revtex4}

\usepackage{graphicx} 
\usepackage{dcolumn}  
\usepackage{rotating} 
\usepackage{lscape}

\usepackage{bm}
\usepackage{url} 
\usepackage{amsmath} 
\usepackage{amssymb} 

\usepackage{verbatim}
\usepackage{multirow}

\graphicspath{{ps}}

\begin{document}

\title{Axion Dark Matter Search around 6.7 $\mu$eV}

\author{S. Lee}\affiliation{Center for Axion and Precision Physics Research, Institute for Basic Science, Daejeon 34051, Republic of Korea}
\author{S. Ahn}\affiliation{Dept. of Physics, Korea Advanced Institute of Science and Technology, Daejeon 34141, Republic of Korea}
\author{J. Choi}\altaffiliation[]{Now at Korea Astronomy and Space Science Institute, Daejeon 34055, Republic of Korea}\affiliation{Center for Axion and Precision Physics Research, Institute for Basic Science, Daejeon 34051, Republic of Korea}
\author{B. R. Ko}\email[Corresponding author~:~]{brko@ibs.re.kr}\affiliation{Center for Axion and Precision Physics Research, Institute for Basic Science, Daejeon 34051, Republic of Korea}
\author{Y. K. Semertzidis}\affiliation{Center for Axion and Precision Physics Research, Institute for Basic Science, Daejeon 34051, Republic of Korea}\affiliation{Dept. of Physics, Korea Advanced Institute of Science and Technology, Daejeon 34141, Republic of Korea}

\begin{abstract}
  An axion dark matter search with the CAPP-8TB haloscope is
  reported.
  Our results are sensitive to axion-photon coupling
  $g_{a\gamma\gamma}$ down to the QCD axion band over the axion mass
  range between 6.62 and 6.82 $\mu$eV at a 90\% confidence level,
  which is the most sensitive result in the mass range to date.
\end{abstract}


\maketitle
\tighten

{\renewcommand{\thefootnote}{\fnsymbol{footnote}}}
\setcounter{footnote}{0}

Precision cosmological measurements strongly favor the standard model
of Big Bang cosmology where about 85\% of the matter in the Universe
is cold dark matter (CDM)~\cite{PLANCK}. However,
CDM itself is beyond the standard model of particle physics (SM),
and accordingly the nature of about 85\% of the matter in the Universe
is still unknown to date. One of the most promising CDM candidates is
the axion~\cite{AXION}, provided its mass is light enough, above 1
$\mu$eV~\cite{CDM_LOW} and below 3 meV~\cite{SN1987}. The axion is the
result of the breakdown of a new symmetry which was proposed by Peccei
and Quinn~\cite{PQ} to solve the strong $CP$ problem in the
SM~\cite{strongCP}. A consequence of the axion production mechanisms
in the early Universe~\cite{CDM_LOW, AXION_PROD2, AXION_PROD3,
  AXION_PROD4} is that the axion mass range is very broad: the range
mentioned above is the optimum for CDM. On the other hand, the open
axion mass range can be much broader according to more recent
works~\cite{AXION_PROD5}.

The axion haloscope search proposed by Sikivie~\cite{sikivie} involves
the resonant conversion of axions to photons in a microwave cavity
permeated by a static magnetic field. The conversion power
corresponding to the axion signal should be enhanced when the axion
mass $m_a$ matches the resonant frequency of the resonator mode $\nu$,
$m_a=h\nu/c^2$. This makes the axion haloscope one of the most
promising methods for axion dark matter searches in the microwave
region which approximately fits the CDM region. The resonated power to
be detected as the axion signal in SI units is as
\begin{equation}
  P^{a\gamma\gamma}_{a}=g^2_{a\gamma\gamma}\frac{\rho_a\hbar^2}{m^2_a c}\omega(2U_M)C Q_L\frac{\beta}{(1+\beta)},
  \label{EQ:PAXION}
\end{equation}
where $g_{a\gamma\gamma}$ is the axion-photon coupling strength.
The two most popular benchmark models are Kim-Shifman-Vainshtein-Zakharov
(KSVZ)~\cite{KSVZ} for axions that couple to beyond the SM heavy
quarks, and Dine-Fischler-Srednicki-Zhitnitskii (DFSZ)~\cite{DFSZ} for
axions that couple to the SM quarks and leptons, at tree levels.
$\rho_a\approx0.45$ GeV/cm$^3$ is the local dark matter density,
$\omega=2\pi\nu$, and $U_M=\frac{1}{2\mu_0}B^2_{\rm
  avg}V\equiv\frac{1}{2\mu_0}\int\vec{B}^2 dV$ is energy stored in a
magnetic field over the resonator volume $V$, where $\vec{B}$ is a
static magnetic field provided by magnets in the axion haloscopes.
The resonator-mode-dependent form factor $C$ whose general definition
can be found in Ref.~\cite{EMFF_BRKO} and loaded quality factor
$Q_L=Q/(1+\beta)$ are also shown in Eq.~(\ref{EQ:PAXION}), where $Q$
is the unloaded quality factor of the resonator mode and $\beta$
denotes the resonator mode coupling to the load.
Assuming the axions have an isothermal distribution, the signal power
given in Eq.~(\ref{EQ:PAXION}) would then distribute over a boosted
Maxwellian shape with an axion rms velocity of about 270 km/s and the
Earth rms velocity of 230 km/s with respect to the galaxy
frame~\cite{AXION_SHAPE}, respectively, which is the model adopted in
this Letter.

Here, we report an axion dark matter search with the
CAPP-8TB haloscope at the {\bf I}nstitute for {\bf B}asic {\bf
  S}cience (IBS) {\bf C}enter for {\bf A}xion and {\bf P}recision {\bf
  P}hysics Research (CAPP)~\cite{CAPP}. The 8TB stands for our
solenoid specifications, the central magnetic field of {\bf 8 T} and
the relatively {\bf B}ig bore of 165 mm.

\begin{figure*}
  \centering
  \includegraphics[width=0.9\textwidth]{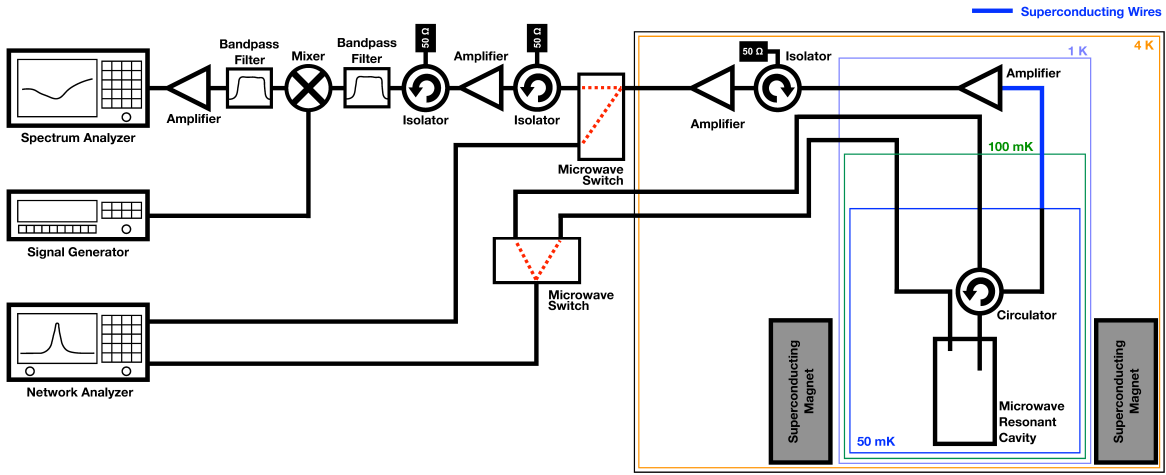}  
  \caption{Overview of the CAPP-8TB axion haloscope.}
  \label{FIG:CAPP-8TB}
\end{figure*}

The CAPP-8TB haloscope has a tunable copper cylindrical cavity as a
resonator, a cryogen-free NbTi superconductor solenoid~\cite{AMI}, and
a typical heterodyne receiver chain equipped with a state-of-the-art
High-Electron-Mobility Transistor (HEMT) LNF-LNC0.6\_2A~\cite{LNF} as
the first amplifier.
The experiment maintains the physical temperature of the cavity at
about 47 mK using a cryogen-free dilution refrigerator
BF-LD400~\cite{BlueFors}.
The details of the CAPP-8TB apparatus will be discussed in the coming
publication~\cite{8TBLong1}. Here we provide an overview shown in
Fig.~\ref{FIG:CAPP-8TB}.

Our cavity is a 3.47 L copper cylinder whose inner diameter and height
are 134 and 246 mm, respectively. The frequency tuning mechanism is
comprised of a fully alumina based tuning rod system, a locomotive
shaft to link the tuning rod axle and a rotational stepper motor.
The motor is operated at room temperature, and the link between
the locomotive shaft and the motor is achieved using Carbon Fiber
Reinforced Polymer (CFRP) tubes. All the rotational axles in the
cavity and the tuning mechanism systems are supported by ball bearings
to minimize friction in the experiment. We found that the locomotive
frequency tuning mechanism, which we developed in this research, does not
introduce any significant heat whenever we tune the resonant
frequencies of the cavity. This eliminates dead time between the
cavity property measurement and power measurement.
The quasi-TM$_{010}$ (QTM$_{010}$) modes of the cavity, which return
the best form factors for axion haloscope searches with cylindrical
geometry, are tuned by rotating the frequency tuning rod system.
A linear stepper motor operated at room temperature and an antenna are
linked with CFRP tubes to adjust $\beta$ between 1.8 and 2.0. A fixed
antenna minimally coupled to the cavity is also employed. The $Q_L$
values of the QTM$_{010}$ modes were obtained over the frequency range
between 1.60 and 1.65 GHz; they were about 30,000. Based on the
$Q_L$ values of the QTM$_{010}$ modes, the discrete frequency step was
chosen to be less than half of the smallest cavity linewidth, i.e.,
20 kHz. The form factors of the QTM$_{010}$ modes were numerically
evaluated with the magnetic field map of the magnet and the electric
field maps of the QTM$_{010}$ modes from
simulations~\cite{CST,COMSOL}, and they are about 0.52 over the
frequency range. The field map of the magnet also returned a $B_{\rm
  avg}$ of 7.3 T over the cavity volume.

Our receiver chain consists of a single data acquisition channel. As
shown in Fig.~\ref{FIG:CAPP-8TB}, power from the cavity goes through
the circulator, the first HEMT amplifier, and the other following
microwave components, and is then measured by a spectrum analyzer at
the end.
Cavity parameters, $\nu$, $Q_L$, and $\beta$ are measured with a
network analyzer by toggling microwave switches.
Having maintained the physical temperature of the first HEMT
amplifier, the gain of the receiver chain $G$ is
\begin{equation}
  G=\frac{P_h-P_c}{k_B\Delta{f}(T_h-T_c)},
  \label{EQ:YGAIN}
\end{equation}
where $P_c$ and $P_h$ are the power measured by the CAPP-8TB haloscope
with cavity temperatures of 50 mK ($T_c$) and 200 mK ($T_h$),
respectively, and $k_B$ is the Boltzmann constant and $\Delta{f}$ is
the resolution bandwidth (RBW). We kept the physical
temperature of the first HEMT amplifier as cold as possible by
attaching it to the 1 K stage~\cite{H_HEMT} as shown in
Fig.~\ref{FIG:CAPP-8TB}, which results in a physical temperature of
about 1.2 K. The measured gains are about 132 dB taking into account
all the attenuation in our receiver chain, over the frequency range
between 1.60 and 1.65 GHz. The noise temperatures of the receiver
chain were obtained using the background parameterization discussed
below.
\begin{figure*}
  \centering
  \includegraphics[width=1.0\textwidth]{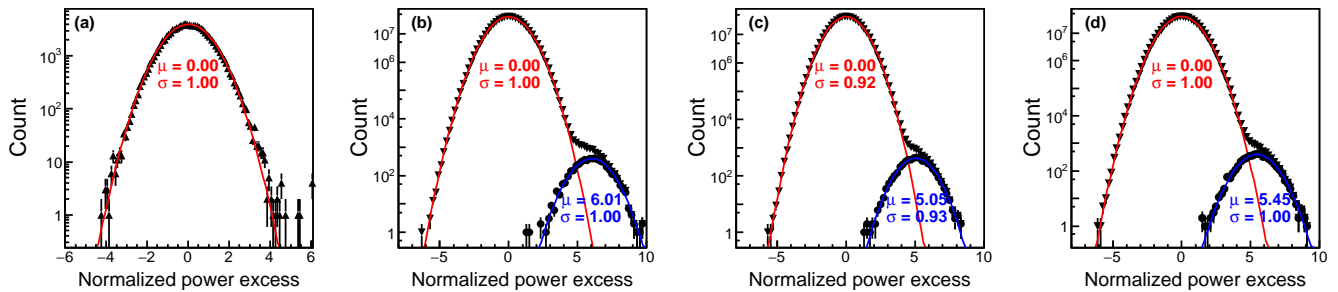}  
  \caption{Triangles in (a) show the distribution of the normalized
    power excess from all the frequency bins in the normalized grand
    power spectrum from the CAPP-8TB experiment, incorporating
    the correlation coefficients between the frequency bins
    participating in the co-adding. Inverted triangles in (b), (c),
    and (d) are the distributions of the normalized power excess from
    all the frequency bins in the normalized grand power spectra from
    the 5,000 simulated axion haloscope search experiments described
    in the text, while the solid circles are those from a particular
    frequency bin where we put simulated axion signals on top the
    CAPP-8TB background. (b) was obtained after subtracting the
    background with perfect fit, while (c) and (d) have a
    five-parameter fit. (d) incorporates the correlations between the
    co-adding frequency bins. Lines are a Gaussian fit resulting in
    $\mu$ (mean) and $\sigma$ (width).} \label{FIG:CAPP-8TB-PULLS}
\end{figure*}

The signal-to-noise ratio (SNR) in this work is defined by the axion
signal power in Eq.~(\ref{EQ:PAXION}) with the coupling
$g_{a\gamma\gamma}\sim4\times g^{\rm KSVZ}_{a\gamma\gamma}$, where
$g^{\rm KSVZ}_{a\gamma\gamma}$ is the standard KSVZ axion coupling
with the neutral heavy quark~\cite{QCD_AXION}, thus is
\begin{equation}
  {\rm SNR}=\frac{P^{{a\gamma\gamma}}_a}{\sigma_{P_n}}=\frac{P^{{a\gamma\gamma}}_a}{P_n}\sqrt{N}\sim\frac{P^{4\times g^{\rm KSVZ}_{a\gamma\gamma}}_a}{k_B b_a T_n}\sqrt{N}
  \label{EQ:SNR}
\end{equation}
according to the radiometer equation~\cite{DICKE}, where $P_n$ and
$\sigma_{P_n}$ are the noise power and its fluctuation,
respectively. The coupling $4\times g^{\rm KSVZ}_{a\gamma\gamma}$
meets the QCD axion~\cite{QCD_AXION}, $N$ is the number of power
spectra, $T_n$ is the system noise temperature, which is a sum of the
noise temperatures of the cavity and the receiver chain, and $b_a$ is
the axion signal window, which is about 5 kHz for the axion masses
considered in this Letter. We acquired data from September 25 to
November 11 in 2019 including a week of system maintenance. A total of
2,501 resonant frequencies were scanned for a search
window of 50 MHz, with frequency steps of 20 kHz. In every resonant
frequency, we collected a total 12,000 power spectra with an RBW of
20 Hz, and averaged them to reach an SNR in Eq.~(\ref{EQ:SNR}) of about
4, which resulted in an SNR of 5 or higher at the end.

Our overall analysis basically follows the previous axion haloscope
search analyses~\cite{haloscope4, HAYSTAC}. We took power spectra over
a frequency span of 60.48 kHz and an RBW of 20 Hz. They were chosen
to maximize the SNR with our frequency steps of 20 kHz and our spectrum
analyzer performance, respectively. The five nonoverlapping  frequency
bins in the individual power spectra are then merged so that the RBW
becomes 100 Hz from 20 Hz. The power spectra with the RBW of 100 Hz went
through a filtering procedure similar to the one developed by the
Haloscope at Yale Sensitive to Axion CDM (HAYSTAC)~\cite{HAYSTAC}, where the
background was subtracted with a five-parameter fit developed by the
Axion Dark Matter eXperiment (ADMX)~\cite{haloscope4}. The efficiency
of the filtering was estimated to be 99.7\% using the simulated axion
experiments described below.
This background parameterization itself was also used to obtain system
noise temperature from the total power and the gain calculated in
Eq.~(\ref{EQ:YGAIN}), as mentioned earlier.
It was about 1 K over the frequency range considered in this
Letter, which is consistent with the first HEMT amplifier
specifications~\cite{LNF}.
The five nonoverlapping frequency bins in each background-subtracted
power spectrum were again merged so that the RBW became 500 Hz from
100 Hz.
We found the signal efficiency with the misalignment between the
axion signal and an RBW of 500 Hz was 99.9\%. All the power spectra
were then combined as a single power spectrum, taking into account
the overlaps among the power spectra. From the combined spectrum, we
constructed our ``grand power spectrum'' by
co-adding~\cite{haloscope4} 10 adjacent 500 Hz power spectral lines
that possess 99.9\% signal power if axions are there, where each power
spectral line was weighted according to the axion signal
shape~\cite{HAYSTAC}.
The grand power spectrum was further normalized by $\sigma_{P_n}$
which was also weighted according to the axion signal
shape~\cite{HAYSTAC}. Using 5,000 simulated axion haloscope search
experiments with CAPP-8TB backgrounds over a search window of 50 MHz
and axion signals at a particular frequency,
the correlation coefficient matrices between the frequency bins
participating in the co-adding were fully constructed and
incorporated into the $\sigma_{P_n}$ for the first time in axion
haloscope searches. The correlations were induced from the background
subtraction~\cite{haloscope4, HAYSTAC, ACTION}.
After employing the correlation coefficient matrices,
Fig.~\ref{FIG:CAPP-8TB-PULLS} (a) shows that the distribution of the
normalized power excess from all frequency bins in the normalized
grand power spectrum from the CAPP-8TB experiment followed the standard
Gaussian distribution, i.e., the width of the triangles is unity.
The simulated experiments also resulted in distributions shown as inverted
triangles, from all the frequency bins, and solid circles from a
particular frequency bin where we put the simulated axion
signals. These are shown as (b), (c), and (d) in
Fig.~\ref{FIG:CAPP-8TB-PULLS}, respectively.
Figure~\ref{FIG:CAPP-8TB-PULLS} (b) was obtained after
the background subtraction using the simulation input background
functions, i.e., perfect fit. (c) and (d) were obtained with a
five-parameter fit. (d) incorporates the correlations between the
co-adding frequency bins, and thus follows the standard Gaussian, while
(c) does not, as reported in Refs.~\cite{haloscope4, HAYSTAC, ACTION}.
The SNR of the CAPP-8TB experiment in each frequency bin corresponding
to the grand power spectrum bin was also calculated. From the
distributions of solid circles in Figs.~\ref{FIG:CAPP-8TB-PULLS} (b),
(c), and (d), the five-parameter fit efficiencies with and without the
correlation coefficient matrices turn out to be about 90.7\% and
84.0\%, respectively. The SNR degradation from the five-parameter fit,
the filtering for the individual power spectrum, and the misalignment
with the RBW of 500 Hz was reflected in the final SNR.
\begin{figure*}
  \centering
  \includegraphics[width=1.0\textwidth]{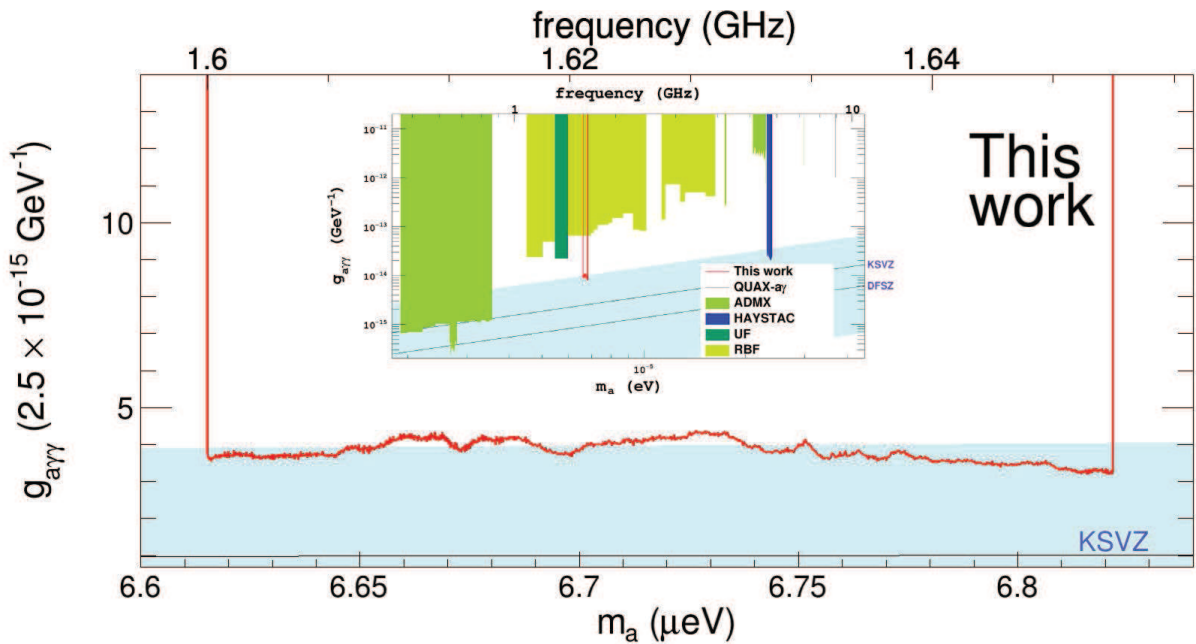}  
  \caption{Red solid line is the excluded parameter space at a 90\%
    confidence level by this work and cyan is the axion model
    band~\cite{QCD_AXION}. The inset shows exclusion limits from other
    axion haloscope searches~\cite{haloscope2, haloscope3, haloscope4,
      haloscope5, ADMX_DFSZ, ADMX_SideCar, HAYSTAC, HAYSTAC2, QUAX} as
    well as that from this work.}
  \label{FIG:CAPP-8TB-LIMIT}
\end{figure*}

We found 36 grand power spectrum bins that exceeded a threshold of
3.718$\sigma_{P_n}$. The 36 excess frequency bins were re-scanned with
sufficient high statistics and no bins survived.
Hence, we set the 90\% upper limits of $g_{a\gamma\gamma}$ for
$6.62<m_a<6.82$ $\mu$eV. Figure~\ref{FIG:CAPP-8TB-LIMIT} shows the
excluded parameter space at a 90\% confidence level from the CAPP-8TB
experiment. Our result is the most sensitive in the relevant axion
mass ranges to date.

In summary, we reported an axion dark matter search using the
CAPP-8TB haloscope. The CAPP-8TB experiment incorporated a developed
locomotive frequency tuning mechanism which generated negligible friction
and also improved the SNR in the axion haloscope search analysis
procedure.
With this improvement, the CAPP-8TB experiment achieved the results
sensitive to axion-photon coupling $g_{a\gamma\gamma}$ down to the QCD
axion band~\cite{QCD_AXION} over the axion mass range between 6.62 and
6.82 $\mu$eV at a 90\% confidence level, which is the most sensitive
result in the mass range to date. We expect to run the CAPP-8TB
haloscope with a quantum-noise-limited superconducting amplifier and
superconducting cavity~\cite{CAPP} to reach the standard KSVZ axion
coupling in the near future.

\acknowledgments
This work was supported by IBS-R017-D1-2020-a00.


\begin{thebibliography}{99}
\bibitem{H_HEMT}
  The HEMT amplifier generates heat power about 14 mW. Applying such
  heat power to the 100 mK stage or below does not let us operate the
  dilution fridge properly.
\end{thebibliography}
\end{document}